\documentclass[conference]{IEEEtran}
\IEEEoverridecommandlockouts
\usepackage{cite}
\usepackage{amsmath,amssymb,amsfonts}
\usepackage{algorithmic}
\usepackage{graphicx}
\usepackage{textcomp}
\usepackage[numbers]{natbib}
\usepackage{xcolor}
\def\BibTeX{{\rm B\kern-.05em{\sc i\kern-.025em b}\kern-.08em
    T\kern-.1667em\lower.7ex\hbox{E}\kern-.125emX}}
\begin{document}

\title{Human-in-the-Loop Systems for Adaptive Learning Using Generative AI}

\author{
    \IEEEauthorblockN{Bhavishya Tarun, Haoze Du, Dinesh Kannan, Dr. Edward F. Gehringer}
    \IEEEauthorblockA{\textit{Department of Computer Science} \\\textit{North Carolina State University} \\
    Raleigh, USA \\
    \{btarun, hdu5, dkannan, efg\}@ncsu.edu}
}

% \thanks{Identify applicable funding agency here. If none, delete this.}
% }

% }

\maketitle

\begin{abstract}
A Human-in-the-Loop (HITL) approach leverages generative AI to enhance personalized learning by directly integrating student feedback into AI-generated solutions. Students critique and modify AI responses using predefined feedback tags, fostering deeper engagement and understanding. This empowers students to actively shape their learning, with AI serving as an adaptive partner. The system uses a tagging technique and prompt engineering to personalize content, informing a Retrieval-Augmented Generation (RAG) system to retrieve relevant educational material and adjust explanations in real time. This builds on existing research in adaptive learning, demonstrating how student-driven feedback loops can modify AI-generated responses for improved student retention and engagement, particularly in STEM education. Preliminary findings from a study with STEM students indicate improved learning outcomes and confidence compared to traditional AI tools. This work highlights AI’s potential to create dynamic, feedback-driven, and personalized learning environments through iterative refinement.
\end{abstract}


\section{Introduction}
The integration of generative AI in education has opened new avenues for personalized learning, offering scalable yet adaptive support for students. However, many existing AI-assisted educational systems operate in a static fashion, delivering content without actively responding to students' evolving needs or encouraging critical engagement. Large Language Models (LLMs), such as GPT-4 (OpenAI, 2023), have shown considerable promise in education, offering the ability to generate contextually appropriate, grammatically fluent, and semantically rich explanations. These models can assist learners by answering questions, explaining complex topics, or even simulating tutoring interactions \cite{2024.EDM-short-papers.49}. However, most existing applications use LLMs in a static fashion: the model provides a one-shot response, and the student either accepts or discards it. This limits the model’s potential to support deeper cognitive engagement, self-regulated learning, or long-term personalization.

In contrast, HITL systems are designed to involve humans not just as passive recipients but as active contributors in the decision-making or learning loop \cite{amershi_power_2014}. In the educational context, HITL strategies empower learners to interact with AI-generated content by critiquing, modifying, or augmenting it in ways that guide future responses. Prior work has applied this paradigm in contexts such as peer review feedback, writing assistance, and automated grading \cite{holstein_designing_2019}, often with teachers or instructors playing the feedback role. However, relatively few studies have focused on student-centered feedback mechanisms that allow learners to shape AI behavior directly and iteratively.

A growing body of research suggests that feedback-rich adaptive systems significantly enhance students’ learning outcomes and metacognitive development \cite{brown_language_2020}. HITL systems have emerged as a promising design paradigm in this context, enabling learners to interact with and influence AI-driven processes \cite{amershi_power_2014}. In education, HITL approaches can serve a dual function: allowing students to reflect on AI-generated content and enabling the system to improve and personalize future outputs. However, most of the existing works on educational AI focus on teacher feedback \cite{jia2024llm}, rubric-guided grading \cite{du2024interactive}, or predefined adaptive pathways \citep{10.1145/3587102.3588799, 10.1145/3626252.3630870}, leaving a gap in research that actively centers student feedback as a mechanism for real-time system adaptation.

To address this challenge, we propose a Human-in-the-Loop framework that integrates generative AI with a student-facing feedback tagging system to foster deeper cognitive engagement and personalized content refinement. Students are encouraged to critique AI-generated solutions using a set of predefined tags (e.g., clarity, correctness, tone), which not only help structure reflection, but also drive the Retrieval-Augmented Generation (RAG) component of the system. These tags are used to retrieve and reformulate more relevant and targeted content in future responses, ensuring that the system remains aligned with the evolving understanding and preferences of the learner. By transforming passive consumption of AI output into an interactive feedback-driven loop, this approach promotes student agency, iterative learning, and scalable personalization.

The proposed system is designed to be domain-agnostic but is evaluated here in the context of STEM education, where the ability to iteratively review, critique and refine problem-solving steps is critical. Our methodology draws on advances in prompt engineering, reinforcement learning, and sentiment-aware AI to ensure that student feedback not only informs but meaningfully alters the system’s behavior in future interactions.

In the remainder of this paper, we first review the theoretical foundations and prior work relevant to Human-in-the-Loop systems, generative AI in education, and feedback-driven personalization in Section 2. Section 3 introduces the design and methodology of our proposed system, detailing its architecture, feedback tagging mechanism, prompt engineering strategy, and integration with Retrieval-Augmented Generation (RAG) for adaptive content delivery. In Section 4, we present the results of our empirical evaluation, including both learning outcome metrics and engagement patterns observed during STEM-based problem solving tasks. Section 5 discusses the implications of these findings for personalized learning, student agency, and the broader application of AI in education. Finally, Section 6 concludes the paper by summarizing key contributions and outlining directions for future research and development.

\section{Related Works}

In this section, we review four key areas of related research: Human-in-the-Loop (HITL) systems in education, the use of large language models (LLMs) for educational purposes, feedback mechanisms and their impact on learning outcomes, and the emerging use of Retrieval-Augmented Generation (RAG) in adaptive systems. This foundation highlights the existing contributions and current limitations in the field that motivate the development of our student-in-the-loop generative feedback system.

\subsection{Human-in-the-Loop (HITL) Learning Systems}
Human-in-the-Loop (HITL) approaches have been widely studied in machine learning and AI system design as a means of improving robustness, fairness, and adaptability through human feedback \cite{amershi_power_2014}. In educational settings, HITL systems are typically used to embed human judgment into automated decision-making or content generation loops. For example, \cite{holstein_designing_2019} proposed a teacher-in-the-loop framework to allow instructors to dynamically monitor and guide intelligent tutoring systems.

However, most of these systems place educators or system designers at the center of the loop. Few efforts have attempted to position students themselves as active participants who critique and shape AI behavior in real time. In contrast, our work shifts the feedback agency to learners, enabling a student-centered HITL interaction that not only promotes engagement but also transforms students into co-constructors of their learning paths. This design also aligns with broader pedagogical shifts toward learner-centered and constructivist approaches in education \cite{learn2000brain}.

\subsection{Generative AI in Education}
Generative models like GPT-3 and GPT-4 \cite{openai_gpt-4_2024} have demonstrated remarkable potential in supporting educational tasks such as question answering, code generation, and essay feedback. While these systems are capable of producing high-quality responses, their lack of interactivity and adaptability limits their effectiveness in fostering personalized learning experiences. In our other recent work \cite{du2024interactive, jia2024llm}, we have been increasingly used LLMs to generate formative feedback on student work, offering scalable and context-aware suggestions that support learning across diverse domains. Some studies have proposed fine-tuning or prompt engineering to align outputs with pedagogical goals \cite{chu_llm_2025}. 

However, a central limitation persists: LLMs are predominantly used in one-way interactions where students passively receive AI-generated output. Without the ability to correct or influence the model’s reasoning, students may fail to develop metacognitive awareness or critical evaluation skills. Although fine-tuning and prompt engineering have been used to improve educational alignment \cite{mazzullo2025fine}, these approaches often lack scalability and do not accommodate individual learner variation in real-time. Our system addresses this gap by combining prompt engineering with student-generated feedback tags to iteratively refine AI responses during the learning process.

\subsection{Feedback Mechanisms in Learning}
Feedback is widely recognized as one of the most effective tools for improving student learning outcomes \cite{hattie2007power}. Effective feedback helps students identify knowledge gaps, reflect on their reasoning, and develop a growth mindset. In the context of intelligent tutoring systems, both formative feedback (provided during learning) and summative feedback (delivered after task completion) have been shown to influence learner motivation and achievement \cite{van2015effects}.

Recent research has emphasized the role of iterative, dialogic feedback — particularly in digital learning environments \cite{nye2014autotutor}. For instance, studies on dialog-based tutoring systems suggest that students benefit more from adaptive follow-up questions and revision prompts than from static explanations. Feedback tagging mechanisms have also been explored as a way to encourage students to evaluate and categorize content \cite{valencia2019effect}, which improves both engagement and comprehension.

\subsection{Retrieval-Augmented Generation (RAG) Systems}
Retrieval-Augmented Generation (RAG) is a relatively recent technique that enhances the factual accuracy and contextual relevance of generative models by coupling them with external knowledge retrieval components \cite{lewis_retrieval-augmented_2020}. In educational applications, RAG enables systems to retrieve course-specific materials, prior explanations, or peer-generated responses to complement the generative process. This can significantly improve alignment with academic goals and reduce hallucination in AI-generated output \cite{izacard2023atlas}.

While most current RAG systems retrieve content based on semantic similarity or keyword matching, few leverage real-time student feedback as a retrieval signal. In our system, tagging data from students is used to augment the retrieval process by indicating which aspects of content were insufficient (e.g., incorrect, unclear) and which should be prioritized (e.g., detail level, tone). This results in an adaptive loop where each iteration becomes more aligned with the student's evolving needs and preferences.

\subsection{Summary of related work}
To summarize, prior research has laid critical groundwork in areas such as HITL system design, LLM-based tutoring, and feedback-informed personalization. However, existing systems often lack mechanisms for dynamic, student-driven adaptation of AI behavior. Most feedback systems are either educator-centered or limited to summative corrections, while LLMs remain largely unresponsive to real-time critique. Furthermore, the integration of RAG systems with learner-generated feedback signals has not yet been fully explored.

Our work addresses these gaps by:
\begin{itemize}
    \item Centering students as the feedback providers in a Human-in-the-Loop architecture.
    \item Using structured tagging not only for evaluation but also as a generative signal.
    \item Combining tagging with RAG to personalize content iteratively and responsively.
\end{itemize}

Through this design, our goal is to create an AI-powered learning system that is adaptive, engaging, and aligned with modern pedagogical principles of student agency and formative feedback.

\section{System Design and Methodology}

\subsection{System Architecture}

\begin{figure}[h]
  \centering
  \includegraphics[width=\linewidth]{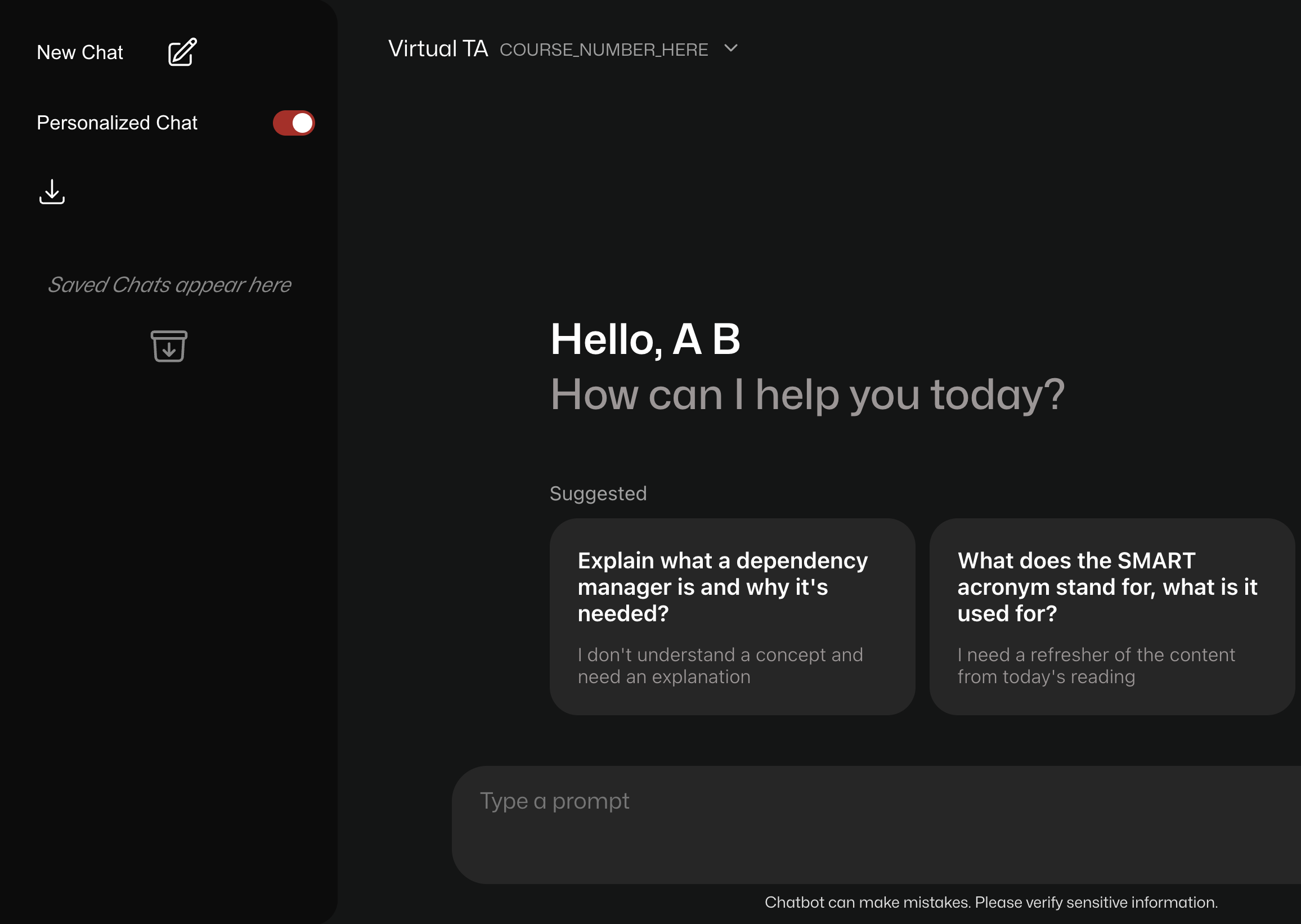}
  \caption{Front‑end chatbot interface.}
  \label{fig:chatbot_ui}
\end{figure}

\begin{figure}[h]
  \centering
  \includegraphics[width=\linewidth]{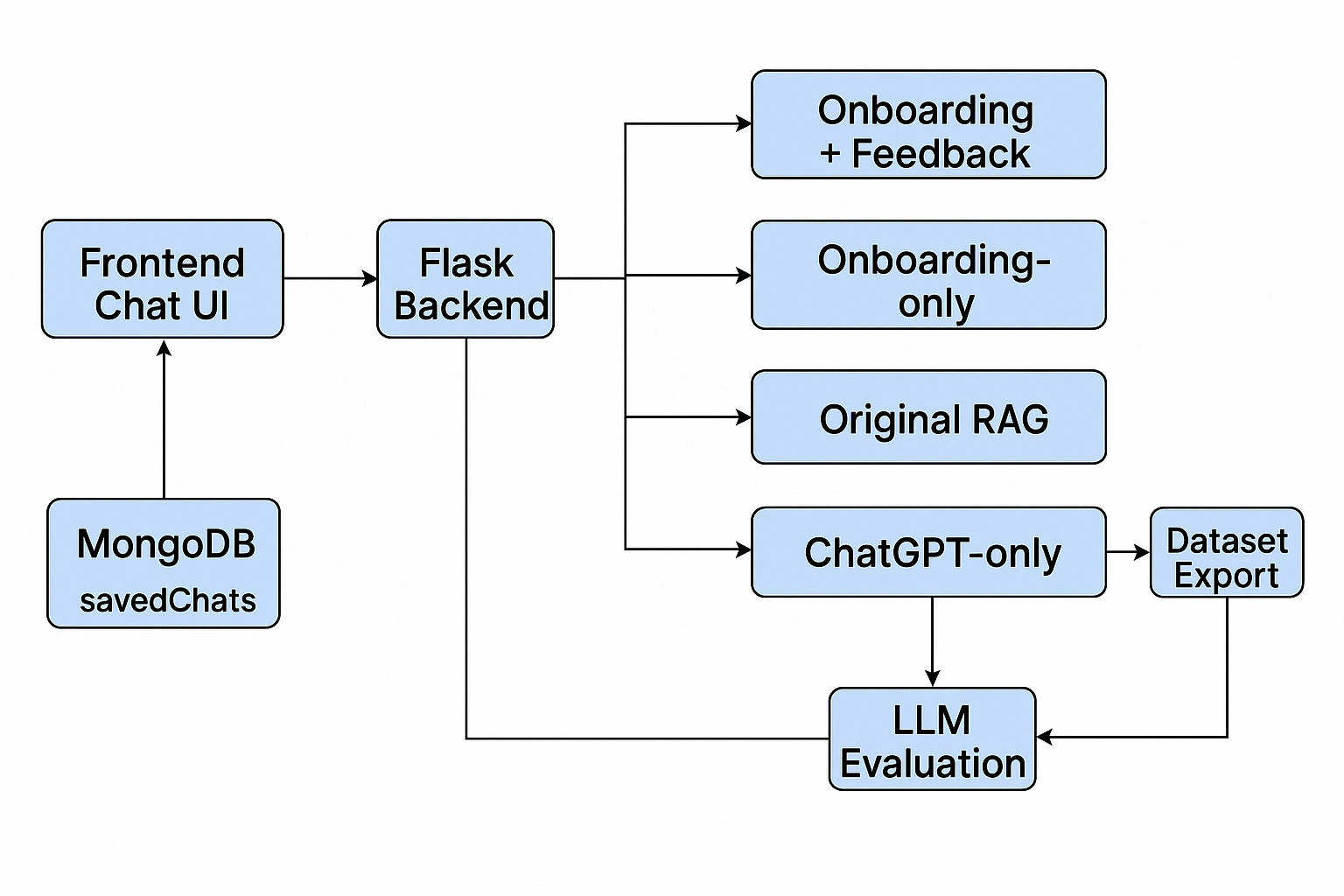}
  \caption{Overall system architecture.}
  \label{fig:architecture}
\end{figure}

As illustrated in Figures \ref{fig:chatbot_ui} and \ref{fig:architecture}, our implementation is a modular web service built on Flask and MongoDB, augmented by a four‑pipeline query manager.  Every incoming question is tagged with a secure base session key, to which we append four fixed suffixes to spawn parallel sessions: \emph{Personalized + Feedback}, \emph{Personalized}, \emph{RAG}, and \emph{LLM}.  All chat data including questions, metadata, and responses are persisted under these session keys in MongoDB, enabling isolated tracking of each pipeline’s output.

\subsection{Onboarding and Feedback Flow}

\begin{figure}[h]
  \centering
  \includegraphics[width=\linewidth]{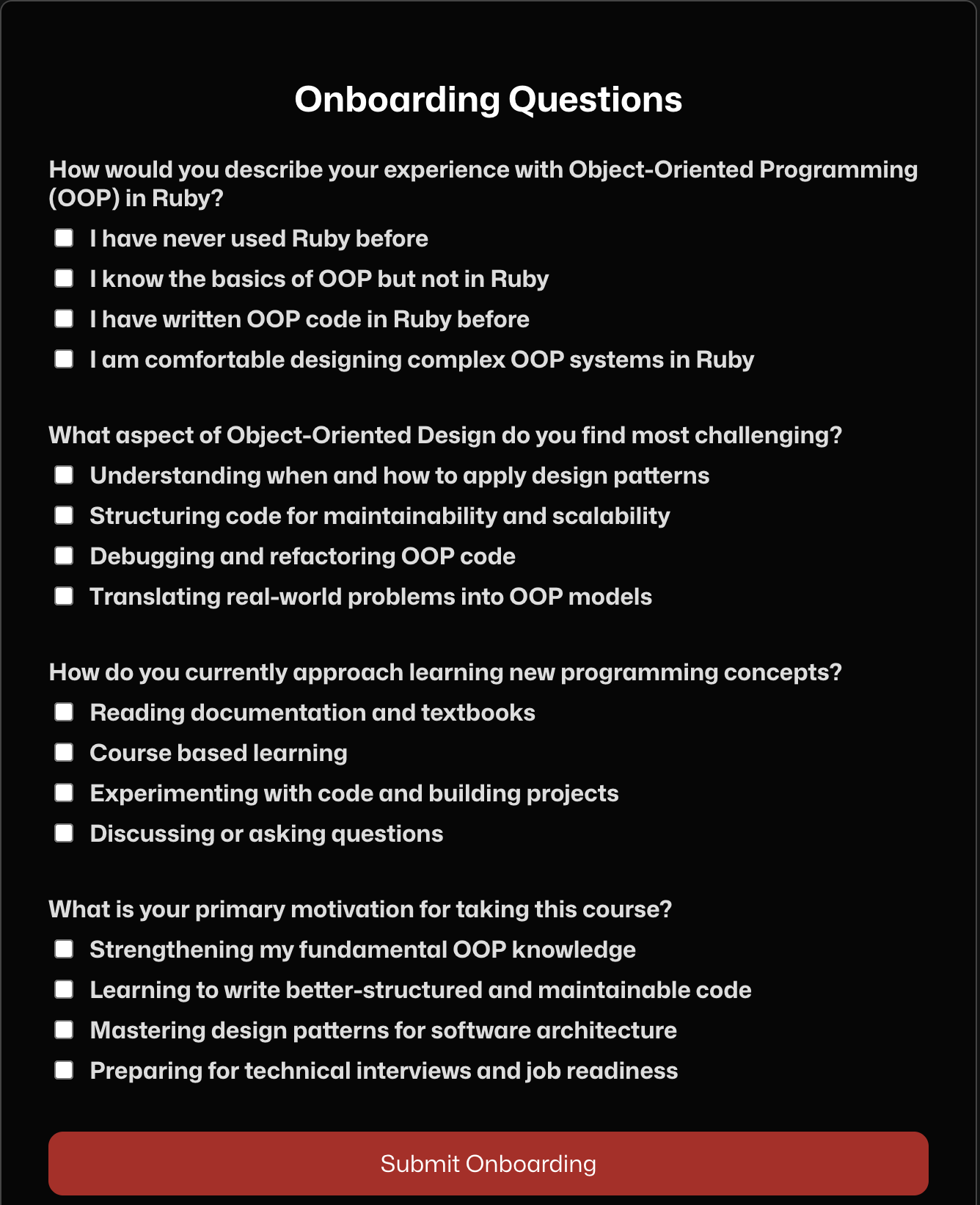}
  \caption{Onboarding questionnaire and in‑chat feedback interface.}
  \label{fig:onboarding_ui}
\end{figure}

Figure~\ref{fig:onboarding_ui} shows the initial pop‑up questionnaire and the inline feedback buttons. At the outset of each session, users complete a concise pop‑up questionnaire capturing their Ruby/OOP experience, preferred learning style, common design challenges, and personal goals. As the conversation proceeds, every bot reply in the \emph{Personalized + Feedback} pipeline includes inline buttons for five predefined tags (“Excellent,” “Very Helpful,” “Average,” “Poor,” “Terrible”).  A user’s selection is immediately recorded and fed back into the next streaming request, allowing the system to adapt its subsequent responses in real time.

\subsection{Four Parallel Query Pipelines}

After onboarding and any live feedback are gathered, we execute four distinct pipelines in parallel.  The \emph{Personalized + Feedback} pipeline merges both static metadata and live tags into a LangChain streaming call.  The \emph{Personalized} pipeline uses only the onboarding metadata.  The \emph{RAG} pipeline retrieves the top ten textbook passages (cosine‐similarity $\geq 0.8$) from our MongoDB vector index, concatenates them with recent chat history and context, and omits any personalization fields.  Finally, the \emph{LLM} pipeline issues a direct call to the language model without retrieval or metadata injection.  Each pipeline’s response is saved under its own session key, and only that response column is populated in our in‑memory dataset buffer.

\subsection{Unified Prompt Template}

To ensure a fair comparison across all four pipelines, we feed each one the same high‑level prompt structure. This template directs the model to answer only from the provided textbook excerpts and conversation history; otherwise instructs the model to state “I’m sorry, I do not know the answer to that question.” It also instructs the LLM to handle greetings, remember user names, include code snippets from the textbook upon request, and incorporate optional fields for user metadata (e.g.\ experience level, learning style) and feedback tags. We define five discrete feedback options—\textbf{Excellent} (clear, insightful, comprehensive), \textbf{Very Helpful} (informative, useful, detailed), \textbf{Average} (adequate, generic, basic), \textbf{Poor} (incomplete, unclear, insufficient), and \textbf{Terrible} (incorrect, irrelevant, unhelpful)—each with a concise interpretation. By toggling only the metadata or feedback variables on or off per pipeline, we isolate the impact of retrieval, static onboarding, and live feedback on final outputs.

\subsection{Dataset Construction and Evaluation}

Upon completing all turns—approximately 200 across eight sessions focused on Object‑Oriented Design—we export the in‑memory log to form our dataset.  Each row represents a single conversational turn, with columns for \emph{Session}, \emph{Personalized + Feedback}, \emph{Personalized}, \emph{RAG}, \emph{LLM}, and \emph{UserPreference} (the JSON‑encoded onboarding answers).  In a post‑hoc phase, we invoke a stringent GPT‑based evaluator to re‑score every response on four metrics—Correctness, Clarity, Readability, and Adaptability.  This process yields both detailed per‑turn scores and aggregated metric tables for each pipeline, forming the basis for our comparative analysis in Section 4.

\section{Results and Evaluation}

\begin{figure*}[!t]
  \centering
  \includegraphics[width=\textwidth]{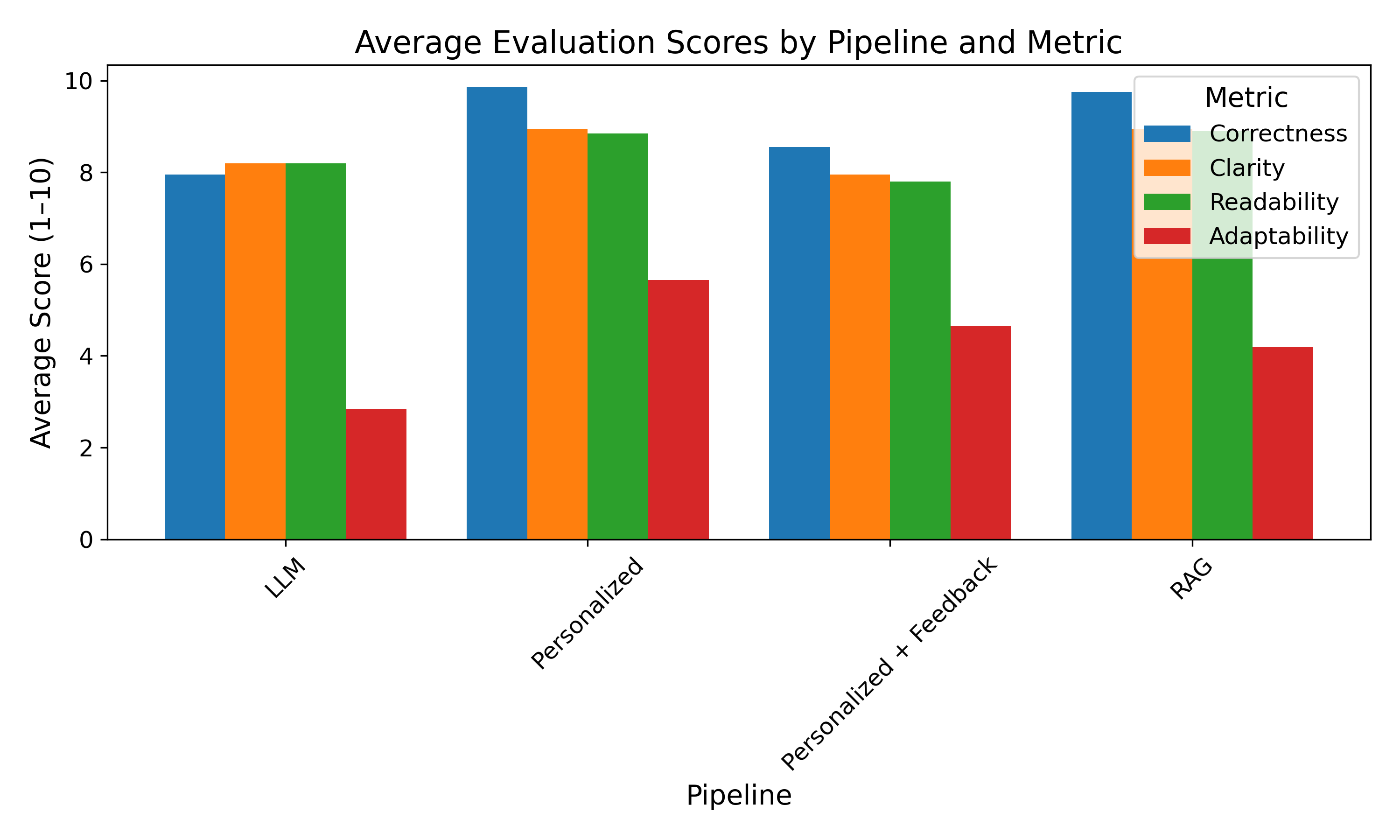}
  \caption{Average evaluation scores by pipeline and metric.}
  \label{fig:pipeline_scores}
\end{figure*}

As shown in Fig.~\ref{fig:pipeline_scores}, we averaged a balanced dataset of approximately 200 conversational turns across eight sessions, with each turn asking an Object‑Oriented Design question drawn from the course textbook. For every turn, each of the four pipelines produced one reply, yielding four versions per question. These responses were then evaluated by a GPT‑based expert, who assigned an integer score from 1 (poor) to 10 (excellent) on four criteria:  
 as Table \ref{tab:metrics_def} shows.

\begin{table*}[t]
    \caption{Metrics utilized to evaluate the quality of generated contents}
    \label{tab:metrics_def}
    \centering
    % \resizebox{\columnwidth}{!}{%
    \begin{tabular}{c|c}
       \hline
       Metric & Definition \\
       \hline
       Correctness & assessing factual accuracy and absence of misleading statements\\
       Clarity  &  how easy the response is to understand and follow\\ 
       Readability & judging conciseness, logical structure, and formatting (including code snippets)\\ 
       Adaptability & evaluating how well the reply reflects the user’s background, preferences, and any live feedback\\
       \hline
    \end{tabular}
    % }
\end{table*}

Table \ref{tab:MLE} presents the average scores for each pipeline. Both the static “Personalized” pipeline and the retrieval‑grounded “RAG” pipeline achieve very high correctness score (9.81 and 9.70 respectively), while the vanilla “LLM” baseline falls behind at 7.89. All pipelines maintain strong clarity and readability (in the 8.0–9.0 range), demonstrating the consistency of our prompt design. Adaptability varies most dramatically: it peaks at 5.43 for the static “Personalized” pipeline, drops to 4.55 when live feedback tags are added, sits at 4.12 for “RAG,” and reaches its lowest point of 2.78 for the untuned “LLM” model. The modest gain from feedback tags suggests that our five feedback tags, applied only at the end of each turn, provide limited real‑time improvement. It is also a case where the extreme options have not been selected. 

\begin{table}[h]
  \caption{Mean evaluation scores (1–10) by pipeline}
  \label{tab:MLE}
  \centering
  \resizebox{\columnwidth}{!}{%
    \begin{tabular}{lcccc}
      \hline
      Pipeline                   & Correctness & Clarity & Readability & Adaptability \\
      \hline
      Personalized + Feedback    & 8.65        & 8.03    & 7.83        & 4.55         \\
      Personalized               & 9.81        & 8.95    & 8.82        & 5.43         \\
      RAG                        & 9.70        & 8.95    & 8.88        & 4.12         \\
      LLM                        & 7.89        & 8.20    & 8.24        & 2.78         \\
      \hline
    \end{tabular}%
  }
\end{table}

Live feedback was used sparingly—only 19 tags were recorded across all turns (see Table~\ref{tab:feedback_counts}).

\begin{table}[h]
  \caption{Feedback tag distribution (turn‑level counts)}
  \label{tab:feedback_counts}
  \centering
    \begin{tabular}{lc}
      \hline
      Tag           & Count \\
      \hline
      Excellent     & 3     \\
      Very Helpful  & 5     \\
      Average       & 6     \\
      Poor          & 3     \\
      Terrible      & 2     \\
      \hline
    \end{tabular}
\end{table}

\begin{figure}[h]
  \centering
  \includegraphics[width=\linewidth]{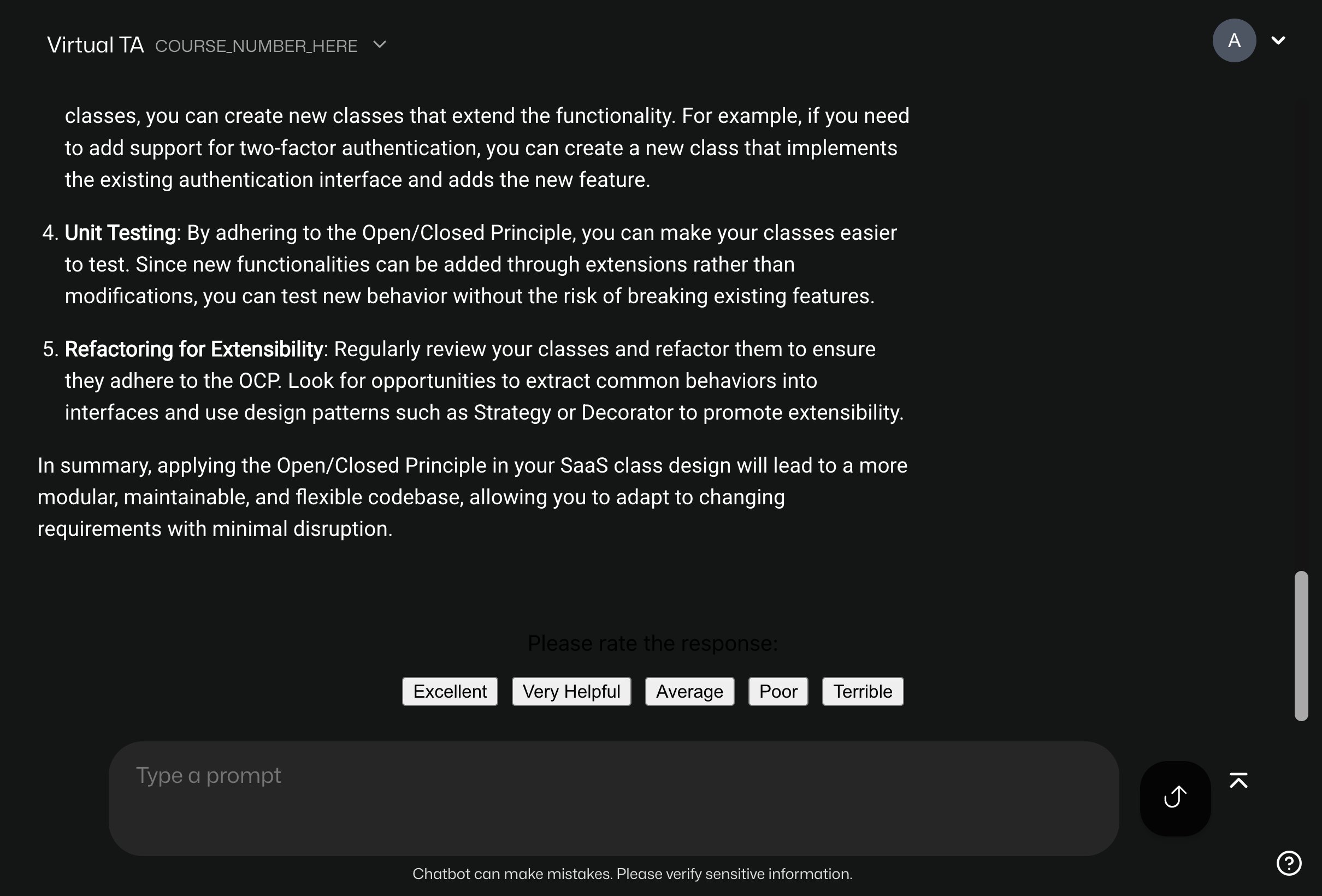}
  \caption{Feedback tags across conversation.}
  \label{fig:feedback}
\end{figure}

We also compared how much each metric varied across pipelines. Adaptability shows the greatest spread (standard deviation $\approx $1.03), whereas Correctness ($\approx $ 0.80), Readability ($\approx $ 0.42), and Clarity ($\approx $0.40) remain relatively stable. This confirms that personalization primarily influences how well replies match individual user needs, while factual and stylistic qualities are managed consistently across all systems.

\section{Discussion}

Our evaluation demonstrates that user profiling yields the most substantial personalization benefit. The static “Personalized” pipeline, which conditions every response on onboarding metadata (experience level, learning style, goals), achieved the highest Adaptability score (5.43). This suggests that stable user characteristics, captured once at the start, provide the model with reliable signals that persist throughout the dialogue.

In contrast, augmenting personalization with live feedback tags resulted in only a modest Adaptability gain (4.55). Two factors likely limit this effect. First, users applied feedback sparingly—only in about 9.5\% of turns—indicating possible tag fatigue or insufficient incentive to engage. Second, our five‐level tag set, applied at the end of each turn, may lack the granularity and immediacy needed to meaningfully steer the model’s next output. Addressing these issues will require both a richer, hierarchical feedback taxonomy and mechanisms to inject feedback more directly into the in‐flight prompt (e.g., mid‐stream updates).

Retrieval grounding remains indispensable for factual accuracy. The RAG pipeline’s Correctness score of 9.70 nearly matches that of the static “Personalized” pipeline, whereas the untuned LLM baseline falls to 7.89. By anchoring responses in domain‑specific textbook passages, the RAG approach effectively mitigates hallucinations and ensures precise, contextually relevant answers.

Clarity and Readability are uniformly strong (all pipelines score between 8.0 and 9.0), confirming that our shared prompt template and formatting conventions reliably elicit coherent, well‑structured responses. By comparison, Adaptability exhibits the greatest variability—standard deviation $\approx $ 1.03 versus $\approx $ 0.80 for Correctness—highlighting it as the most sensitive indicator of how well each pipeline aligns with individual learner needs.

Taken together, these findings support a complementary architecture: static onboarding metadata delivers consistent personalization, retrieval ensures factual precision, and live feedback—while promising—must be rethought to achieve higher engagement and finer control.

\section{Conclusion}

In this paper, we presented a modular Human‑in‑the‑Loop framework that orchestrates four parallel AI pipelines—\emph{Personalized + Feedback}, \emph{Personalized}, \emph{RAG}, and a plain \emph{LLM} baseline—to support Object‑Oriented Design tutoring. Our systematic comparison over roughly 200 conversational turns revealed two key insights: first, static onboarding metadata alone delivers the strongest personalization benefit, boosting Adaptability to 5.43; second, retrieval grounding in a domain‑specific textbook achieves near‑perfect Correctness ($\approx$ 9.7), while the untuned LLM, despite clear and readable output, lags on both accuracy and personalization. Live feedback tagging, though conceptually promising, yielded only marginal Adaptability gains (4.55) under our five‑tag schema and turn-end injection, underscoring the need for more engaging and fine‑grained mechanisms.

\subsection{Limitations}  
Our evaluation is confined to a single textbook domain and relies on a GPT‑based evaluator for all scoring, which may introduce both topical and evaluator bias. User engagement with live feedback was low ($\approx$ 9.5\% of turns), limiting our ability to fully assess its impact. Moreover, our coarse, end‑of‑turn tagging schema and the absence of a human rater constrain the precision and generalizability of our findings.

\subsection{Future Work}  
Future work will focus on expanding the feedback schema, integrating tags mid‑prompt, enriching user profiles with additional signals (e.g., past performance, sentiment), and migrating to a multi‑annotator evaluation platform to improve both engagement and evaluation reliability. We also plan to broaden our study across multiple subjects and incorporate real‑time analytics dashboards, thereby paving the way toward truly adaptive, student‑centered learning environments.

Additionally, while the current system demonstrates promising results in terms of personalized learning and engagement in controlled environments, future work will involve deploying the system in real classroom settings to collect large-scale statistical data. By embedding the feedback-tagging interface into authentic STEM courses, we aim to observe how students interact with generative AI under realistic learning conditions and over longer time spans. This will allow us to analyze longitudinal patterns in student behavior, engagement trends, and the evolving nature of feedback tags in response to course content. Moreover, the collected data will facilitate robust statistical evaluations, such as correlation analyses between tag usage and learning gains, as well as cluster analyses to identify learner profiles. These insights will support further refinement of the tagging schema, retrieval strategy, and personalization mechanisms, ultimately improving the effectiveness and generalizability of the Human-in-the-Loop framework across diverse educational contexts.

\bibliographystyle{IEEEtran}
\bibliography{references}

% Generated by IEEEtran.bst, version: 1.14 (2015/08/26)
\begin{thebibliography}{10}
\providecommand{\url}[1]{#1}
\csname url@samestyle\endcsname
\providecommand{\newblock}{\relax}
\providecommand{\bibinfo}[2]{#2}
\providecommand{\BIBentrySTDinterwordspacing}{\spaceskip=0pt\relax}
\providecommand{\BIBentryALTinterwordstretchfactor}{4}
\providecommand{\BIBentryALTinterwordspacing}{\spaceskip=\fontdimen2\font plus
\BIBentryALTinterwordstretchfactor\fontdimen3\font minus \fontdimen4\font\relax}
\providecommand{\BIBforeignlanguage}[2]{{%
\expandafter\ifx\csname l@#1\endcsname\relax
\typeout{** WARNING: IEEEtran.bst: No hyphenation pattern has been}%
\typeout{** loaded for the language `#1'. Using the pattern for}%
\typeout{** the default language instead.}%
\else
\language=\csname l@#1\endcsname
\fi
#2}}
\providecommand{\BIBdecl}{\relax}
\BIBdecl

\bibitem{2024.EDM-short-papers.49}
Q.~Jia, J.~Cui, R.~Xi, C.~Liu, P.~Rashid, R.~Li, and E.~Gehringer, ``On assessing the faithfulness of llm-generated feedback on student assignments,'' in \emph{Proceedings of the 17th International Conference on Educational Data Mining}, B.~PaaÃŸen and C.~D. Epp, Eds.\hskip 1em plus 0.5em minus 0.4em\relax Atlanta, Georgia, USA: International Educational Data Mining Society, July 2024, pp. 491--499.

\bibitem{amershi_power_2014}
S.~Amershi, M.~Cakmak, W.~B. Knox, and T.~Kulesza, ``\BIBforeignlanguage{en}{Power to the {People}: {The} {Role} of {Humans} in {Interactive} {Machine} {Learning}},'' \emph{\BIBforeignlanguage{en}{AI Magazine}}, vol.~35, no.~4, pp. 105--120, Dec. 2014, number: 4.

\bibitem{holstein_designing_2019}
K.~Holstein, B.~M. McLaren, and V.~Aleven, ``\BIBforeignlanguage{en}{Designing for {Complementarity}: {Teacher} and {Student} {Needs} for {Orchestration} {Support} in {AI}-{Enhanced} {Classrooms}},'' in \emph{\BIBforeignlanguage{en}{Artificial {Intelligence} in {Education}}}, S.~Isotani, E.~Millán, A.~Ogan, P.~Hastings, B.~McLaren, and R.~Luckin, Eds.\hskip 1em plus 0.5em minus 0.4em\relax Cham: Springer International Publishing, 2019, pp. 157--171.

\bibitem{brown_language_2020}
T.~Brown, B.~Mann, N.~Ryder, M.~Subbiah, J.~D. Kaplan, P.~Dhariwal, A.~Neelakantan, P.~Shyam, G.~Sastry, A.~Askell, S.~Agarwal, A.~Herbert-Voss, G.~Krueger, T.~Henighan, R.~Child, A.~Ramesh, D.~Ziegler, J.~Wu, C.~Winter, C.~Hesse, M.~Chen, E.~Sigler, M.~Litwin, S.~Gray, B.~Chess, J.~Clark, C.~Berner, S.~McCandlish, A.~Radford, I.~Sutskever, and D.~Amodei, ``Language {Models} are {Few}-{Shot} {Learners},'' in \emph{Advances in {Neural} {Information} {Processing} {Systems}}, vol.~33.\hskip 1em plus 0.5em minus 0.4em\relax Curran Associates, Inc., 2020, pp. 1877--1901.

\bibitem{jia2024llm}
Q.~Jia, J.~Cui, H.~Du, P.~Rashid, R.~Xi, R.~Li, and E.~Gehringer, ``{LLM}-generated feedback in real classes and beyond: Perspectives from students and instructors,'' in \emph{Proceedings of the 17th International Conference on Educational Data Mining}, 2024, pp. 862--867.

\bibitem{du2024interactive}
H.~Du, P.~Rashid, Q.~Jia, and E.~Gehringer, ``Interactive rubric generator for instructor's assessment using prompt engineering and large language models,'' in \emph{2024 IEEE Frontiers in Education Conference (FIE)}.\hskip 1em plus 0.5em minus 0.4em\relax IEEE, 2024, pp. 1--9.

\bibitem{10.1145/3587102.3588799}
\BIBentryALTinterwordspacing
J.~Cui, R.~Zhang, R.~Li, Y.~Song, F.~Zhou, and E.~Gehringer, ``Correlating students' class performance based on github metrics: A statistical study,'' in \emph{Proceedings of the 2023 Conference on Innovation and Technology in Computer Science Education V. 1}, ser. ITiCSE 2023.\hskip 1em plus 0.5em minus 0.4em\relax New York, NY, USA: Association for Computing Machinery, 2023, p. 526–532. [Online]. Available: \url{https://doi-org.prox.lib.ncsu.edu/10.1145/3587102.3588799}
\BIBentrySTDinterwordspacing

\bibitem{10.1145/3626252.3630870}
\BIBentryALTinterwordspacing
J.~Cui, R.~Zhang, R.~Li, F.~Zhou, Y.~Song, and E.~Gehringer, ``How pre-class programming experience influences students' contribution to their team project: A statistical study,'' in \emph{Proceedings of the 55th ACM Technical Symposium on Computer Science Education V. 1}, ser. SIGCSE 2024.\hskip 1em plus 0.5em minus 0.4em\relax New York, NY, USA: Association for Computing Machinery, 2024, p. 255–261. [Online]. Available: \url{https://doi-org.prox.lib.ncsu.edu/10.1145/3626252.3630870}
\BIBentrySTDinterwordspacing

\bibitem{learn2000brain}
H.~P. Learn, ``Brain, mind, experience, and school,'' \emph{Committee on Developments in the Science of Learning}, pp. 14--15, 2000.

\bibitem{openai_gpt-4_2024}
\BIBentryALTinterwordspacing
OpenAI, J.~Achiam, and {et al.}, ``{GPT}-4 {Technical} {Report},'' Mar. 2024, arXiv:2303.08774 [cs]. [Online]. Available: \url{http://arxiv.org/abs/2303.08774}
\BIBentrySTDinterwordspacing

\bibitem{chu_llm_2025}
\BIBentryALTinterwordspacing
Z.~Chu, S.~Wang, J.~Xie, T.~Zhu, Y.~Yan, J.~Ye, A.~Zhong, X.~Hu, J.~Liang, P.~S. Yu, and Q.~Wen, ``{LLM} {Agents} for {Education}: {Advances} and {Applications},'' Mar. 2025, arXiv:2503.11733 [cs]. [Online]. Available: \url{http://arxiv.org/abs/2503.11733}
\BIBentrySTDinterwordspacing

\bibitem{mazzullo2025fine}
E.~Mazzullo, O.~Bulut, C.~Walsh, G.~Sitarenios, and A.~MacIntosh, ``Fine-tuning gpt-3.5-turbo for automatic feedback generation,'' 2025.

\bibitem{hattie2007power}
J.~Hattie and H.~Timperley, ``The power of feedback,'' \emph{Review of educational research}, vol.~77, no.~1, pp. 81--112, 2007.

\bibitem{van2015effects}
F.~M. Van~der Kleij, R.~C. Feskens, and T.~J. Eggen, ``Effects of feedback in a computer-based learning environment on students’ learning outcomes: A meta-analysis,'' \emph{Review of educational research}, vol.~85, no.~4, pp. 475--511, 2015.

\bibitem{nye2014autotutor}
B.~D. Nye, A.~C. Graesser, and X.~Hu, ``Autotutor and family: A review of 17 years of natural language tutoring,'' \emph{International Journal of Artificial Intelligence in Education}, vol.~24, pp. 427--469, 2014.

\bibitem{valencia2019effect}
N.~Valencia-Vallejo, O.~L{\'o}pez-Vargas, and L.~Sanabria-Rodr{\'\i}guez, ``Effect of a metacognitive scaffolding on self-efficacy, metacognition, and achievement in e-learning environments.'' \emph{Knowledge Management \& E-Learning}, vol.~11, no.~1, pp. 1--19, 2019.

\bibitem{lewis_retrieval-augmented_2020}
P.~Lewis, E.~Perez, A.~Piktus, F.~Petroni, V.~Karpukhin, N.~Goyal, H.~Küttler, M.~Lewis, W.-t. Yih, T.~Rocktäschel, S.~Riedel, and D.~Kiela, ``Retrieval-{Augmented} {Generation} for {Knowledge}-{Intensive} {NLP} {Tasks},'' in \emph{Advances in {Neural} {Information} {Processing} {Systems}}, vol.~33.\hskip 1em plus 0.5em minus 0.4em\relax Curran Associates, Inc., 2020, pp. 9459--9474.

\bibitem{izacard2023atlas}
G.~Izacard, P.~Lewis, M.~Lomeli, L.~Hosseini, F.~Petroni, T.~Schick, J.~Dwivedi-Yu, A.~Joulin, S.~Riedel, and E.~Grave, ``Atlas: Few-shot learning with retrieval augmented language models,'' \emph{Journal of Machine Learning Research}, vol.~24, no. 251, pp. 1--43, 2023.

\end{thebibliography}
\end{document}